\documentclass[a4paper,fleqn]{cas-sc}

\usepackage[authoryear,longnamesfirst]{natbib}

\def\tsc#1{\csdef{#1}{\textsc{\lowercase{#1}}\xspace}}
\tsc{WGM}
\tsc{QE}

\usepackage{bm}
\usepackage{xcolor,ifthen}
\usepackage{amsmath}
\usepackage{float}
\restylefloat{figure}
\usepackage{setspace}
\usepackage{graphicx}
\usepackage{subcaption}

\newcommand{\Pec}{\mathrm{Pe}}

\begin{document}

\let\WriteBookmarks\relax
\def\floatpagepagefraction{1}
\def\textpagefraction{.001}

\shorttitle{Evidence of an inertialess Kapitza instability due to viscosity stratification}    

\shortauthors{Gundavarapu, Dhas, Roy}  

\title [mode = title]{Evidence of an inertialess Kapitza instability due to viscosity stratification}  

%

\author[1]{Shravya Gundavarapu}

\affiliation[1]{organization={Department of Applied Mechanics and Biomedical Engineering, Indian Institute of Technology Madras},
            city={Chennai},
            postcode={600036}, 
            country={India}}


\author[2]{Darish Jeswin Dhas}

\author[1]{Anubhab Roy}
\cormark[1]

\affiliation[2]{organization={Department of Engineering and Architecture, University of Udine},
            city={Udine},
            postcode={33100}, 
            country={Italy}}

\cortext[1]{anubhab@iitm.ac.in}


\begin{abstract}
The classical Kapitza instability of a gravity-driven falling film requires finite inertia to operate. We show that a surface-mode instability can arise in the complete absence of inertia when the film possesses a continuous viscosity stratification — a feature relevant to particle-laden films with shear-induced migration, thermally stratified coatings, and concentration-graded flows. The viscosity field, prescribed as a linear profile across the film thickness, evolves through an advection--diffusion equation characterized by a P\'{e}clet number.  Using long-wave asymptotics and Chebyshev spectral computations, we solve the coupled eigenvalue problem for the perturbation streamfunction and viscosity fields and demonstrate that viscosity stratification destabilizes the surface mode in the zero-inertia (Stokes) limit. The instability is confined to a finite window of P\'{e}clet numbers. Increasing the stratification parameter lowers the critical P\'{e}clet number, broadens the range of unstable wavenumbers, and increases the growth rate. The instability mechanism is traced to the phase relationship between perturbation vorticity and the interface displacement: viscosity stratification shifts the vorticity to a lagging configuration, which reinforces interface deformation, following the framework of Hinch (1984). The mechanism bears a structural resemblance to the surfactant-driven Marangoni instability in creeping two-layer flows, extending this class of scalar-mediated, inertialess instabilities to bulk viscosity stratification.
\end{abstract}


\begin{highlights}

\item Viscosity stratification drives a surface-mode instability without inertia.
\item Instability exists within a finite P\'{e}clet number window.
\item Instability due to vorticity mismatch via Hinch's mechanism.
\item Mechanism is structurally analogous to surfactant-driven Marangoni instability.

\end{highlights}

\begin{keywords}
 \sep Viscosity stratification \sep Falling film \sep Stokes flow \sep Inertialess instability \sep Vorticity-interface phase mechanism
\end{keywords}

\maketitle

\section{Introduction}
Shallow free-surface flows driven by gravity, even in the absence of microstructural effects, are known to exhibit two distinct classes of instabilities: the surface mode and the shear mode \citep{floryan1987}. The surface mode, first analyzed by Benjamin \citep{benjamin1957} and Yih \citep{yih1963} using linear stability theory, arises at long wavelengths with a critical threshold at $O(1)$ Reynolds number. A defining feature of this mode is the emergence of waves that propagate at nearly twice the mean velocity of the film. In contrast, the shear mode appears at short wavelengths and typically at high Reynolds numbers, with waves traveling slower than the mean velocity \citep{floryan1987}. Unlike the surface mode, where disturbance amplitudes are concentrated near the free surface, shear-mode perturbations are characterized by peak intensities localized near the substrate \citep{chin1986}. Kelly et al.\ \citep{kelly1989} elucidated the mechanism of the surface-mode instability from the viewpoint of vorticity, showing that the phase difference between perturbation vorticity and interface displacement determines whether the flow amplifies or decays - a framework originally proposed by Hinch \citep{hinch1984} for interfacial instabilities in two-layer shear flows. These classical results establish the fundamental instability mechanisms governing falling films of clear, homogeneous fluids.

Viscosity stratification is ubiquitous in both natural and industrial fluid flows, arising from variations in temperature, composition, concentration, or pressure. Its influence on hydrodynamic stability has been studied extensively across a range of flow configurations, and the resulting body of work can be broadly organized according to whether the stratification is discontinuous (sharp interface between layers of different viscosity) or continuous (smooth viscosity variation), and whether the geometry involves wall-bounded channel flows, free-surface or interfacial flows, or gravity-driven configurations.

The foundational theoretical development by Yih \citep{yih1967} showed that two superposed layers of viscous fluid with different viscosities can become unstable even at arbitrarily small Reynolds numbers. The instability arises from a discontinuity in the velocity gradient across the viscosity interface: the jump in viscosity forces a jump in the shear rate to maintain continuity of tangential stress, and this mismatch excites interfacial modes distinct from classical inertial instabilities. Yih demonstrated this for both plane Couette and plane Poiseuille flows, establishing viscosity stratification as a destabilizing mechanism that operates independently of inertia and can alter the stability landscape of otherwise unconditionally stable flows, such as plane Couette flow. Kao \citep{kao1965a,kao1965b,kao1968} extended the analysis to two-layer flows down an inclined plane, investigating the roles of viscosity ratio, density ratio, and depth ratio on the surface and interfacial modes, and showing that interfacial instability arises when the upper layer is more viscous or less dense than the lower layer. Yiantsios \& Higgins \citep{yiantsios1988} carried out stability analyses of two-layer Poiseuille flows and demonstrated how viscosity ratios and interfacial dynamics influence the growth of disturbances. Mohammadi \& Smits \citep{mohammadi2017} studied two-layer Couette flows and mapped the stability boundaries as functions of viscosity ratio, thickness ratio, interfacial tension, and density ratio, demonstrating that viscosity contrast alone is sufficient to destabilize otherwise stable configurations. Chen \citep{chen1993} examined the formation and evolution of waves at the interface of two thin viscous films in the gravity-driven, low-Reynolds-number regime, showing that the interfacial mode can grow even when inertial effects are negligible, provided sufficient viscosity contrast exists between the layers, and Loewenherz \& Lawrence \citep{loewenherz1989} investigated the effect of viscosity stratification on the stability of free-surface flows in the low-Reynolds-number limit, establishing that viscosity-layered configurations can sustain long-wave instabilities independent of inertia.

Craik \citep{craik1969} was among the first to recognize that a continuous viscosity stratification can play a fundamentally different role from a discrete interface. Analyzing plane Couette flow with a smoothly varying viscosity profile, he showed that when viscosity varies continuously, the stability properties are governed by the behavior of the viscosity profile at the critical layer - the location where the base-flow velocity equals the wave speed. Specifically, he demonstrated that the sign and magnitude of the viscosity gradient at the critical layer determine whether the flow is stabilized or destabilized, an effect that has no counterpart in the discontinuous case. Craik \& Smith \citep{craik1968} extended this analysis to free-surface flows with continuous viscosity stratification. Drazin \citep{drazin1962} provided early theoretical results on the stability of parallel flows with variable density and viscosity. Wall \& Wilson \citep{wall1996} examined channel flows with temperature-dependent viscosity and decomposed the effect of non-uniform viscosity into three mechanisms: bulk viscosity modification, velocity-profile reshaping, and thin-layer formation, each influencing stability differently. Sameen \& Govindarajan \citep{sameen2007} analyzed the effect of wall heating on channel flow instability, showing that heating or cooling one wall can dramatically alter the critical Reynolds number through the temperature-dependent viscosity profile. Goussis \& Kelly \citep{goussis1985,goussis1987} used long-wave theory to show that in falling films with exponentially temperature-dependent viscosity, heating (reducing viscosity near the wall) lowers the critical Reynolds number, while cooling acts to stabilize.

 A distinct line of investigation concerns miscible fluids with different viscosities, where the interface is not sharp but diffuse and evolves through molecular diffusion. Govindarajan \citep{govindarajan2004} studied the effect of miscibility on the linear instability of two-fluid channel flow, showing that the overlap region where viscosity varies smoothly introduces new modes of instability absent in the immiscible limit. Talon \& Meiburg \citep{talon2011} investigated the linear stability of miscible, viscosity-layered Poiseuille flow and made the striking observation that in the Stokes flow regime, diffusion has a destabilizing effect analogous to that of inertia in finite-Reynolds-number flows. They identified four types of instability depending on the interface location: two interfacial modes with large growth rates and two bulk modes that grow more slowly. Their work demonstrated that viscosity stratification combined with scalar diffusion can trigger instabilities even in the complete absence of inertia. Usha et al.\ \citep{usha2013} extended the analysis of miscible two-fluid flows to inclined geometries. Lin \citep{lin1946} provided early general results on the stability of viscous parallel flows, and the review by Govindarajan \& Sahu \citep{govindarajan2014} offers a comprehensive survey of instabilities in viscosity-stratified flows across these different configurations.

Early observations by Timberlake \& Morris \citep{timberlake2005} experimentally demonstrated that in neutrally buoyant suspensions flowing down inclined planes, shear-induced particle migration concentrates particles near the free surface, establishing significant viscosity gradients that alter wave dynamics even at vanishing Reynolds numbers. Dhas \& Roy \citep{dhas2022} performed a linear stability analysis of this system in the low-Reynolds-number, finite-P\'{e}clet-number regime, incorporating shear-induced migration, particle-phase normal stresses, and a suspension-balance transport equation, and showed that the resulting viscosity gradients can lower stability thresholds even in the Stokes limit. Their full-suspension model, however, involves several coupled ingredients: the base-state velocity profile is modified by the concentration-dependent viscosity (it is no longer the Nusselt parabola), normal stresses contribute additional forcing to the momentum balance, and the concentration field obeys a nonlinear transport equation distinct from simple advection-diffusion. It is therefore unclear whether viscosity stratification alone is sufficient to trigger the instability, or whether the additional couplings inherent in the suspension model are essential. The question is sharpened by the related study of Dhas \& Roy \citep{dhasroy2022prf} on colloidal falling films, where Brownian diffusion equilibrates the particle concentration to leading order, producing a uniform viscosity increase that stabilizes both the surface and shear modes. That result shows that a particle-induced viscosity change need not be destabilizing; the outcome depends on whether the viscosity field develops a coherent phase-shifted perturbation - a condition that, as we show below, requires intermediate P\'{e}clet numbers and a non-zero base-state viscosity gradient.
 
The present study directly addresses this question. We consider a gravity-driven falling film with a prescribed linear viscosity profile that evolves through a simple advection-diffusion equation 
characterized by a P\'{e}clet number $\Pec$, retaining the classical Nusselt base-state velocity profile. By intentionally decoupling the viscosity field from any underlying concentration or migration dynamics, we isolate viscosity stratification as the sole destabilizing agent. The analysis reveals that this minimal ingredient set is indeed sufficient: a surface-mode instability exists within a finite $\Pec$ window even in the complete absence of inertia. The instability mechanism is elucidated through the vorticity-interface phase framework of Hinch \cite{hinch1984} and Kelly et al.\ \cite{kelly1989}, and is shown to be structurally analogous to the surfactant-driven Marangoni instability in creeping two-layer flows 
\citep{frenkel2002,wei2005}. The paper is organized as follows: the mathematical formulation is introduced in~\S2; the linear stability analysis, including both long-wave and finite-wavenumber results, is presented in~\S3; the mechanism of instability is discussed in~\S4; and conclusions are drawn in~\S5.

\section{Physical problem and governing equations}

\begin{figure}[ht]
        \centering
        \includegraphics[width = 0.5\linewidth]{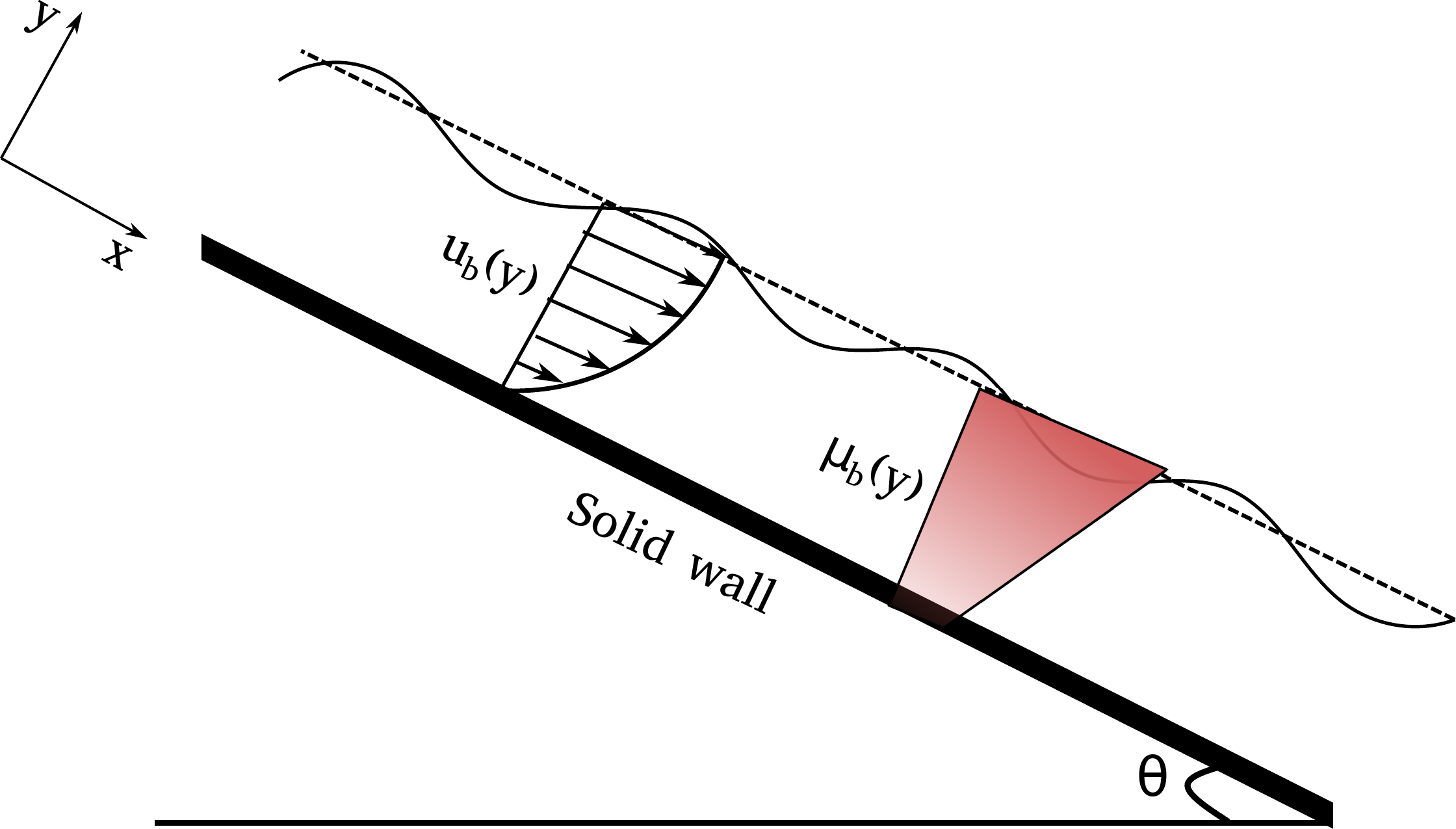}
        \caption{ Schematic of a gravity-driven, viscosity-stratified falling film on an inclined plane with the base flow characterized by a unidirectional velocity profile $u_b(y)$ and a prescribed wall-normal viscosity stratification $\mu_b(y)$.}
        \label{fig:1}
\end{figure} 

We consider a two-dimensional, viscosity-stratified, gravity-driven film flowing down a rigid plane inclined at an angle $\theta$ to the horizontal, as shown in figure \ref{fig:1}. The film has an instantaneous free-surface height $h(x,t)$, and the coordinate system is chosen such that $x$ and $y$ denote the streamwise and wall-normal directions, respectively. In the limit of negligible fluid inertia, the flow is governed by the incompressible Stokes equations
\begin{gather}
\nabla \cdot \boldsymbol{u} = 0, \label{eq:cont_dim} \\
\nabla \cdot \boldsymbol{T} + \rho \boldsymbol{g} = 0, \label{eq:momentum_dim}
\end{gather}
where $\boldsymbol{u} = (u,v)$ is the velocity field, $\rho$ is the fluid density, and $\boldsymbol{g}$ is the gravitational acceleration. The stress tensor $\boldsymbol{T}$ is given by
\begin{equation}
\boldsymbol{T} = -P \boldsymbol{I} + \mu\left( \nabla \boldsymbol{u} + \nabla \boldsymbol{u}^{T} \right),
\end{equation}
where $P$ denotes the pressure and $\mu$ denotes the spatially varying dynamic viscosity.
The flow is subject to no-slip and no-penetration conditions at the rigid substrate ($y=0$),
\begin{equation}
\boldsymbol{u} = \boldsymbol{0}.
\end{equation}
The balance of normal and tangential stresses at the deformable free surface $y=h(x,t)$ is expressed compactly as
\begin{equation}
\boldsymbol{T}\cdot \boldsymbol{n} = \sigma \boldsymbol{n}(\nabla \cdot \boldsymbol{n}),
\end{equation}
where $\sigma$ is the surface tension and $\boldsymbol{n}$ is the outward unit normal vector,
\begin{equation}
\boldsymbol{n} = \frac{(-\partial_x h,\,1)}{\sqrt{1+(\partial_x h)^2}}.
\end{equation}
The free surface evolves according to the kinematic condition
\begin{equation}
\frac{\partial h}{\partial t} + u\frac{\partial h}{\partial x} = v,
\label{eq:kinematic_bc}
\end{equation}
which ensures that the interface is advected by the fluid motion.

The viscosity field is assumed to be advected by the flow and undergoes molecular diffusion, described by an advection-diffusion equation
\begin{equation}
\frac{\partial \mu}{\partial t} + \boldsymbol{u}\cdot \nabla \mu
= \nabla \cdot (D_s \nabla \mu),
\label{eq:mu_transport_dim}
\end{equation}
where $D_s$ is the effective viscosity diffusion coefficient, which we assume to be a constant. Equation \eqref{eq:mu_transport_dim} is supplemented by no-flux boundary conditions at both the solid substrate and the free surface,
\begin{equation}
\boldsymbol{n}\cdot \nabla \mu = 0.
\label{eq:mu_noflux}
\end{equation}
It is important to note that, in the above description, the viscosity field is allowed to evolve both spatially and temporally, while remaining agnostic as to the mechanisms driving these variations.

We nondimensionalize lengths using the mean film thickness $h_0$, velocities using the characteristic falling-film velocity $u_0 = \rho g h_0^2 \sin\theta/3\mu_f$, and the capillary pressure scaling ($\sigma/h_0$) is adopted because, in the zero-inertia limit, the balance between viscous and surface-tension stresses governs the free-surface dynamics. Viscosity is scaled by $\mu_f$ so that $\mu_b = 1$ in the unstratified limit. This nondimensionalization introduces two governing parameters: capillary number $Ca=\mu_f u_0/\sigma$, which measures the ratio of viscous to capillary stresses and governs the resistance of the free surface to deformation, and the Péclet number $\Pec=u_0 h_0/D_s$, which compares advective transport of viscosity to its diffusive relaxation across the film thickness. Subsequently, the non-dimensional system of equations is written as

\begin{gather}
\frac{\partial u}{\partial x} + \frac{\partial v}{\partial y} = 0, \label{eqn:nondim_contI} \\ 
0 = - \frac{\partial P}{\partial x} + \frac{\partial}{\partial x} \left( \mu \frac{\partial u}{\partial x} \right) + \frac{\partial}{\partial y} \left( \mu \frac{\partial u}{\partial y} \right) 
+ \frac{\partial \mu}{\partial y} \frac{\partial v}{\partial x} - \frac{\partial \mu}{\partial x} \frac{\partial v}{\partial y}  + 3,  \\
0 = - \frac{\partial P}{\partial y} + \frac{\partial}{\partial x} \left( \mu \frac{\partial v}{\partial x} \right) + \frac{\partial}{\partial y} \left( \mu \frac{\partial v}{\partial y} \right) 
+ \frac{\partial \mu}{\partial x} \frac{\partial u}{\partial y} - \frac{\partial \mu}{\partial y} \frac{\partial u}{\partial x} - 3 \cot \theta, \\
Pe \left( \frac{\partial \mu}{\partial t} + u \frac{\partial \mu}{\partial x} + v \frac{\partial \mu}{\partial y} \right) = \frac{\partial^2 \mu}{\partial x^2} + \frac{\partial^2 \mu}{\partial y^2}.  \label{eqn:nondim_PhiI}
\end{gather}
The boundary conditions at $y=0$ are
\begin{gather}
u = v = 0, \hspace{4mm} \frac{\partial \mu}{\partial y} = 0, \label{eqn:nondim_bc_0}
\end{gather}
and at the free surface $y=h$, the boundary conditions comprise the normal stress balance,
\begin{equation}
P = \frac{2 \mu}{ \left[ 1 + \left( \frac{\partial h}{\partial x} \right) ^2 \right]} \left[ \left( \frac{\partial u}{\partial x} \left( \frac{\partial h}{\partial x} \right) ^2 - \frac{\partial v}{\partial x} \frac{\partial h}{\partial x} \right) - \frac{\partial u}{\partial y} \frac{\partial h}{\partial x}  + \frac{\partial v}{\partial y} \right] - \frac{ Ca \; \frac{\partial ^2 h}{\partial x ^2}}{\left[ 1 + \left( \frac{\partial h}{\partial x} \right) ^2 \right] ^{3/2}},  \\
\end{equation}
the tangential stress balance,
\begin{equation}
0 = 4 \mu \frac{\partial u}{\partial x} \frac{\partial h}{\partial x} - \mu \left( 1 - \left( \frac{\partial h}{\partial x} \right) ^2 \right) \left( \frac{\partial u}{\partial y} + \frac{\partial v}{\partial x} \right),  \\
\end{equation}
the no-flux condition for the viscosity field,
\begin{equation}
\frac{\partial h}{\partial x} \frac{\partial \mu}{\partial x} - \frac{\partial \mu}{\partial y} = 0, \label{eqn:nondim_bc_h} \hspace{4mm}
\end{equation}
and the kinematic condition,
\begin{equation}
\frac{\partial h}{\partial t} + u \frac{\partial h}{\partial x} = v. 
\end{equation}
For brevity, non-dimensional quantities are denoted by the same symbols as their dimensional counterparts.


\section{Linear stability analysis}

For the linear stability analysis, we assume the base state to be a steady, unidirectional flow that remains unaffected by the viscosity stratification and follows classical Nusselt profile written as
\begin{equation}
u_b(y) = 3\left(y - \frac{y^2}{2}\right).
\end{equation}
We assume the base-state viscosity as a vertically stratified linear profile,
\begin{equation} \label{eqn:mu_b}
\mu_b(y) = 1 + \alpha\,(y-0.5).
\end{equation}
Here, $\alpha$ denotes the strength of viscosity stratification across
the film thickness, with the admissible range being $|\alpha| < 2$ to ensure
$\mu_b > 0$ everywhere in the domain. In all results presented below, we consider $\alpha > 0$, 
corresponding to viscosity increasing toward the free surface.
 
We note that the Nusselt profile $u_b(y)$ is not the exact steady solution of the momentum equation when $\mu_b$ varies with $y$ since the self-consistent velocity profile satisfying $\mathcal{D}[\mu_b\,\mathcal{D} u_b] + 3 = 0$ differs from the Nusselt parabola by an $O(\alpha)$ correction. However, the present choice is made deliberately. By holding the base flow fixed, the analysis ensures that the destabilization solely arises from perturbation-level coupling between the viscosity and velocity fields, and not from modifications to the mean shear. 



We perform a normal mode analysis by decomposing each variable 
as the sum of its base-state value and a small-amplitude 
perturbation with wavenumber $k$ and complex wave speed 
$c = c_r + i c_i$. The sign of $c_i$ determines the stability 
of the mode: $c_i > 0$ corresponds to temporal growth 
(instability), while $c_i < 0$ indicates decay. Following this, we decompose the physical variables in the system into their base-state and perturbed components as 
\begin{equation}
X(x,y,t) = X_b(y) + \hat{X}(y)\,\mathrm{e}^{i k (x - c t)},
\end{equation}
where $X_b(y)$ refers to the base flow variables and $\hat{X}(y)$ refers to the infinitesimally small amplitude of the disturbances. We introduce the perturbation streamfunction $\hat{\psi}$ through $\hat{u} = \mathcal{D}\hat{\psi}$ and 
$\hat{v} = -ik\hat{\psi}$, which automatically satisfies the 
linearised continuity equation. Upon linearisation and 
elimination of the pressure perturbation, the governing 
equations for $\hat{\psi}$ and the viscosity perturbation 
$\hat{\mu}$ reduce to the following coupled system. Here and throughout, $\mathcal{D} \equiv \mathrm{d}/\mathrm{d}y$
denotes the wall-normal derivative operator, and primes indicate derivatives of base-state quantities with respect to $
y$. 
\begin{gather} 
\left\lbrace - 2 \mu_b' (\mathcal{D}^2 - k^2) \mathcal{D} - \mu_b (\mathcal{D}^2 - k^2)^2 - \mu_b '' (\mathcal{D}^2 + k^2) \right\rbrace \hat{\psi} = (\mathcal{D}^2 + k^2) (u_b' \hat{\mu}), \label{eqn:mom_vis} \\ i k \Pec\left[ (u_b - c) \hat{\mu} - \mu_b' \hat{\psi} \right] = (\mathcal{D}^2 - k^2) \hat{\mu}. \label{eq:par1_vis} 
\end{gather}
For the linear base-state viscosity profile adopted in 
equation~\eqref{eqn:mu_b}, $\mu''_b = 0$.

The instability mechanism, as we will describe in section \ref{sec:mechanism}, is driven by the source term $\mu_b'\hat{\psi}$ in the viscosity transport equation and the feedback through $(\mathcal{D}^2+k^2)(u_b'\hat{\mu})$ in the momentum equation. This term operates at the same perturbation order independently of the base-flow correction from viscosity stratification for the moderate values of $\alpha$ considered here ($\alpha \leqslant 0.5$), thus further justifying our choice of base-state velocity field.


The perturbation equations are subject to no-slip and no-flux conditions at the rigid substrate $y=0$,
\begin{gather}
\hat{\psi} = 0, \label{eq:wall_vel_vis} \\
\mathcal{D}\hat{\psi} = 0, \\
\mathcal{D}\hat{\mu} = 0, 
\end{gather}
and to linearised stress and flux conditions at the free 
surface $y=1$. The tangential stress condition gives
\begin{equation}\label{eqn:tang_bc_vis}
\left\lbrace \mu_b \left( \mathcal{D}^2 + k^2 \right) - 
\frac{3}{(c - u_b(1))} \right\rbrace \hat{\psi} = 0,
\end{equation}
The normal stress condition yields
\begin{equation}\label{eqn:norm_bc_vis}
\left\lbrace \mu_b' (\mathcal{D}^2 + k^2) + \mu_b 
(\mathcal{D}^2 - 3 k^2) \mathcal{D} - \left(3 \cot \theta 
+ \frac{k^2}{\mathrm{Ca}} \right) \frac{i k}{(c- u_b(1))} 
\right\rbrace \hat{\psi} + \mathcal{D}(u_b' \hat{\mu}) = 0,
\end{equation}
and the no-flux condition for the viscosity perturbation is
\begin{equation}\label{eq:flux_bc1_vis}
\mathcal{D} \hat{\mu} = 0.
\end{equation}
Together, the governing equations~\eqref{eqn:mom_vis}-\eqref{eq:par1_vis} and boundary conditions~\eqref{eq:wall_vel_vis}-\eqref{eq:flux_bc1_vis} constitute a coupled eigenvalue problem for the perturbation fields $(\hat{\psi}, \hat{\mu})$, which we study in the following sections.


\subsection{Long-wavelength approximation}\label{sec:longwave}

It is well known that gravity-driven falling films are susceptible to long-wavelength disturbances. Accordingly, to examine the influence of viscosity stratification, 
We first consider the long-wave limit $k \ll 1$, in which the disturbance wavelength is large compared to the film thickness,  and expand the perturbation fields and the complex wave speed as
\begin{equation} \label{eqn:longwave_expansion}
\begin{aligned}
\hat{\mu} &= \hat{\mu}_0 + k\,\hat{\mu}_1 + \cdots, \\
\hat{\psi} &= \hat{\psi}_0 + k\,\hat{\psi}_1 + \cdots, \\
c &= c_0 + k\,c_1 + \cdots.
\end{aligned}
\end{equation}
Substitution of the long-wave expansions \eqref{eqn:longwave_expansion} into the linearized governing equations and boundary conditions, followed by collection of terms at $\mathcal{O}(1)$, yields
\begin{gather}
\mathcal{D}^2\!\left( \mu_b \mathcal{D}^2 \hat{\psi}_0 + u_b' \hat{\mu}_0 \right) = 0, \label{eq:psi0_eqn} \\
\mathcal{D}^2 \hat{\mu}_0 = 0, \label{eq:mu0_eqn}
\end{gather}
with the associated boundary conditions at the free surface ($y=1$) 
\begin{gather}
\left\{ \mu_b \mathcal{D}^2 - \frac{3}{c_0 - u_b(1)} \right\} \hat{\psi}_0 = 0, \label{eq:psi0_bc1} \\
\mathcal{D}\!\left( \mu_b \mathcal{D}^2 \hat{\psi}_0 + u_b' \hat{\mu}_0 \right) = 0, \label{eq:psi0_bc2} \\
\mathcal{D} \hat{\mu}_0 = 0, \label{eq:mu0_bc1}
\end{gather}
while at the rigid substrate ($y=0$),
\begin{gather}
\hat{\psi}_0 = \mathcal{D} \hat{\psi}_0 = 0, \label{eq:psi0_wall} \\
\mathcal{D} \hat{\mu}_0 = 0. \label{eq:mu0_wall}
\end{gather}

Solving the above system yields a quadratic dispersion relation for the leading-order wave speed as
\begin{equation} \label{eqn:c0}
a_1 c_0^2 + a_2 c_0 + a_3 = 0,
\end{equation}
where the coefficients $a_1$, $a_2$, and $a_3$ are integral expressions determined by the base-state velocity and viscosity profiles. Their explicit forms are provided in Appendix~\ref{appA}.

\begin{figure}[h]
        \centering
        \includegraphics[width = 0.7\linewidth]{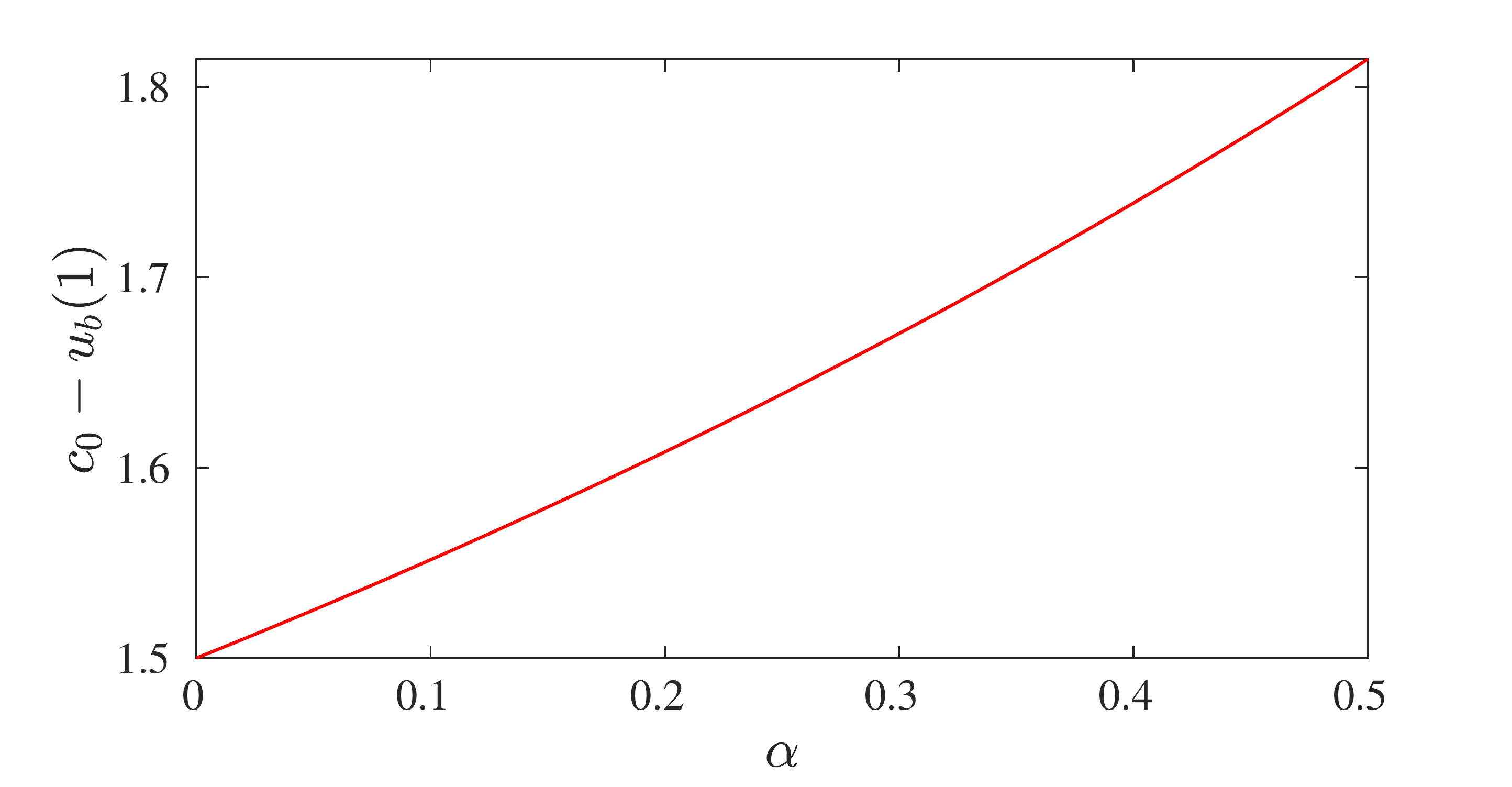}
        \caption{Variation of the Doppler-shifted wave speed $(c_0 - u_b(1))$ of the long-wave surface mode as a function of the stratification parameter $\alpha$, for $\theta = 45^\circ$.}
        \label{fig:2}
\end{figure}

Equation \eqref{eqn:c0} yields two distinct real roots. One root corresponds to the classical long-wave surface mode, modified by the presence of viscosity stratification. The second root arises from the viscosity field, governed by an advection-diffusion equation. Figure~\ref{fig:2} shows the dependence of the Doppler-shifted phase speed of the long-wave surface mode on the viscosity stratification parameter $\alpha$. 
To determine the stability of the long-wave surface mode, we proceed to the next order in the expansion.

Collecting terms at $\mathcal{O}(k)$, the perturbation equations for the streamfunction and viscosity corrections take the form
\begin{gather}
\mathcal{D}^2\!\left( \mu_b \mathcal{D}^2 \hat{\psi}_1 + u_b' \hat{\mu}_1 \right) = 0, \label{eq:psi1_eqn} \\
i\,\Pec\left[(u_b-c_0)\hat{\mu}_0 - \mu_b' \hat{\psi}_0\right] = \mathcal{D}^2 \hat{\mu}_1. \label{eq:mu1_eqn}
\end{gather}
The corresponding boundary conditions at the solid surface ($y=0$) remain homogeneous,
\begin{gather}
\hat{\psi}_1 = \mathcal{D}\hat{\psi}_1 = 0, \\
\mathcal{D}\hat{\mu}_1 = 0,
\end{gather}
while at the free surface ($y=1$), the $\mathcal{O}(k)$ corrections introduce forcing terms proportional to the leading-order solution,
\begin{gather}
\mathcal{D}\!\left( \mu_b \mathcal{D}^2 \hat{\psi}_1 + u_b' \hat{\mu}_1 \right)
= \frac{i\,\mathcal{F}}{c_0-u_b(1)}\,\hat{\psi}_0, \label{eq:psi1_bc1} \\
\mu_b \mathcal{D}^2 \hat{\psi}_1 - \frac{3}{c_0-u_b(1)} \hat{\psi}_1
- \frac{3c_1}{(c_0-u_b(1))^2}\hat{\psi}_0 = 0, \label{eq:psi1_bc2} \\
\mathcal{D}\hat{\mu}_1 = 0, 
\label{eq:mu1_bc2} 
\end{gather}
where $\mathcal{F} = 3\cot\theta + k^2/Ca$.

This $\mathcal{O}(k)$ system \eqref{eq:psi1_eqn}–\eqref{eq:mu1_bc2} constitutes a forced boundary-value problem, which yields a correction $c_1$ to the wave speed. After algebraic manipulation, the resulting expression for $c_1$ can be written in the form
\begin{equation}
c_1 =  i\,c_{1i}, \label{eqn:c1_form}  
\end{equation}
where the imaginary part $c_{1i}$ denotes the temporal growth or decay of the disturbance.
In the interest of brevity, the expression for $c_1$ involving integral operators is in Appendix~\ref{appA} - equation \eqref{eqn:c1}. Evaluating these integrals reveals that within a finite window of $\alpha$ and Péclet numbers $Pe$, $c_{1i} > 0$, confirming the existence of an instability at $\mathcal{O}(k)$. The dependence on $Pe$ enters through the $\mathcal{O}(k)$ viscosity correction $\hat{\mu}_1$, which is governed by the advection–diffusion balance in equation \ref{eq:mu1_eqn}. 

\subsection{Finite wave number analysis}
\begin{figure}[ht]
        \centering
        \includegraphics[width = 0.8\linewidth]{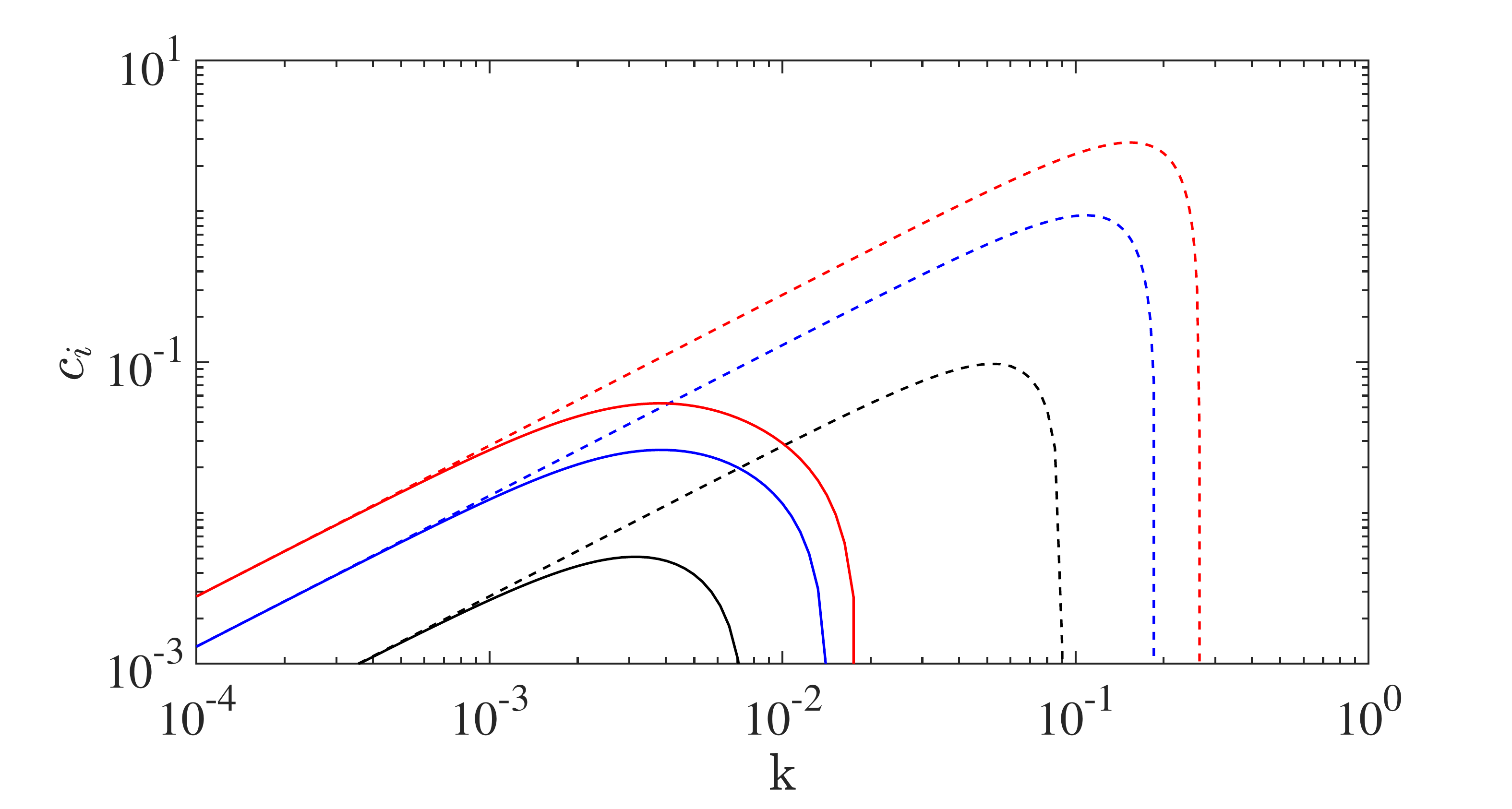}
        \caption{Dispersion relation showing the growth rate $c_i$ versus wavenumber $k$ for different values of the viscosity stratification parameter $\alpha$ at $\Pec=1000$. The black, blue, and red curves correspond to $\alpha=0.1$, $0.3$, and $0.5$, respectively. Solid and dashed lines represent numerical and analytical results, respectively.} 
        \label{fig:3a}
\end{figure}

To investigate the stability characteristics beyond the long-wave limit, we next proceed to solve the full linearized system of equations~\eqref{eqn:mom_vis}-\eqref{eq:flux_bc1_vis} without invoking the small wave number approximation. Towards this, we solve the eigenvalue problem numerically using the Chebyshev spectral collocation method. We map the wall-normal coordinate $y \in [0, 1]$ into $N$ Chebyshev collocation points, and construct differentiation matrices \cite{trefethen2000spectral}. Subsequently, we impose boundary conditions by replacing the corresponding rows of the discretized operators, and solve the resulting generalized eigenvalue problem $\mathcal{A}x = c\,\mathcal{B}x$ using the QZ algorithm implemented in MATLAB.


Figure~\ref{fig:3a} shows the growth rate $c_i$ as a function of the wavenumber $k$ for different values of the viscosity stratification parameter $\alpha$ and $\mathrm{Pe} = 1000$. Unless otherwise specified, we fix $\theta = 45^\circ$ and $\mathrm{Ca} = 0.001$ throughout the paper. For all stratified cases ($\alpha > 0$), the growth rate is positive over a finite range of small wavenumbers, confirming the existence of an instability. The growth rate increases with $\alpha$, and the range of unstable wavenumbers broadens, indicating that stronger stratification enhances the amplitude and broadens its spectral range. The long‑wave asymptotics accurately capture the $c_i \sim \mathcal{O}(k)$ scaling at small $k$, with excellent agreement between the numerical and analytical results. The growth rate attains a maximum at a finite wavenumber and then decreases, eventually becoming negative at sufficiently large $k$. This restabilization at short wavelengths is due to the stabilizing effect of surface tension. This behavior confirms that the instability is long-wavelength and confined to sufficiently small values of $k$.

\begin{figure}[ht]
        \centering
        \includegraphics[width = 1\linewidth]{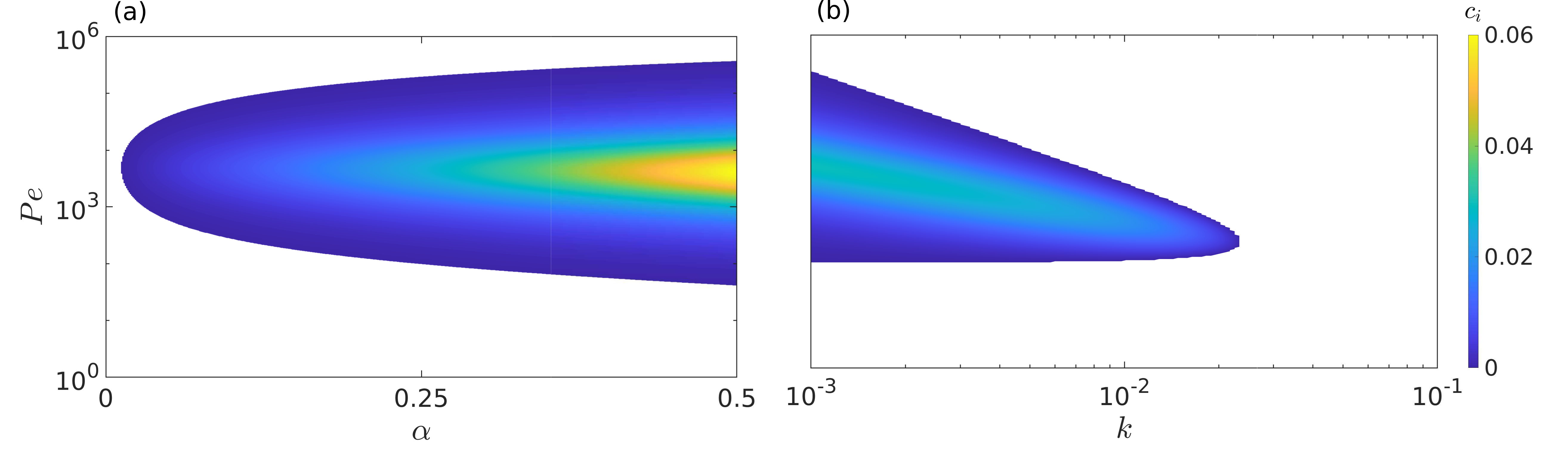}
        \caption{  Contours of the growth rate $\mathrm{Im}(c)$ for long-wave disturbances. (a) $(\alpha,\Pec)$ plane at $k=0.001$. (b) $(k,\Pec)$ plane at $\alpha=0.3$. }
        \label{fig:3}
\end{figure}

To study the combined influence of the viscosity stratification and the scalar transport, we first examine the stability characteristics in the $(\alpha,\Pec)$ parameter plane at fixed wavenumber $k = 0.001$. Figure~\ref{fig:3}(a) presents a contour map of the growth rate, $\mathrm{Im}(c)$, showing a well-defined unstable region bounded by a closed curve. The unstable region forms a lobe-like structure in parameter space, showing that instability occurs only within a finite range of stratification strength and P\'{e}clet number. For very small $\alpha$, the flow remains stable at low $\Pec$, and instability arises only when the P\'{e}clet number exceeds a finite critical value. This behavior shows the stabilizing role of diffusion, when $\Pec$ is small, scalar diffusion smooths viscosity 
perturbations before they can interact with the base shear. As $\Pec$ increases, advective transport sustains viscosity 
perturbations against diffusive smoothing, and the growth 
rate increases. With increasing $\Pec$, the unstable region expands, and the growth rate increases, attaining a maximum at intermediate $\Pec$. However, instability does not persist indefinitely as $\Pec$ increases. Beyond a certain threshold, further increase in the P\'{e}clet number leads to restabilization, and the unstable region gradually shrinks and ultimately vanishes at sufficiently large $\Pec$.

Following the analysis in the $(\alpha,\Pec)$ plane, we examine the stability characteristics in the $(k,\Pec)$ plane at fixed $\alpha = 0.3$ (see Figure \ref{fig:3}(b)). The stability characteristics in the $(k, \Pec)$ plane at fixed $\alpha = 0.3$ (figure~\ref{fig:3}b) complement the 
preceding analysis. Instability is confined to a wedge-shaped band of wavenumbers that exists only over a restricted range of P\'{e}clet numbers, bounded both from below and above. The maximum growth rate occurs at intermediate $\Pec$ within this band, consistent with the 
non-monotonic $\Pec$ dependence identified in the $(\alpha, \Pec)$ plane. The physical mechanism underlying this instability is discussed in \S\ref{sec:mechanism-Pe}.

\begin{figure}[h]
        \centering
        \includegraphics[width = 0.8\linewidth]{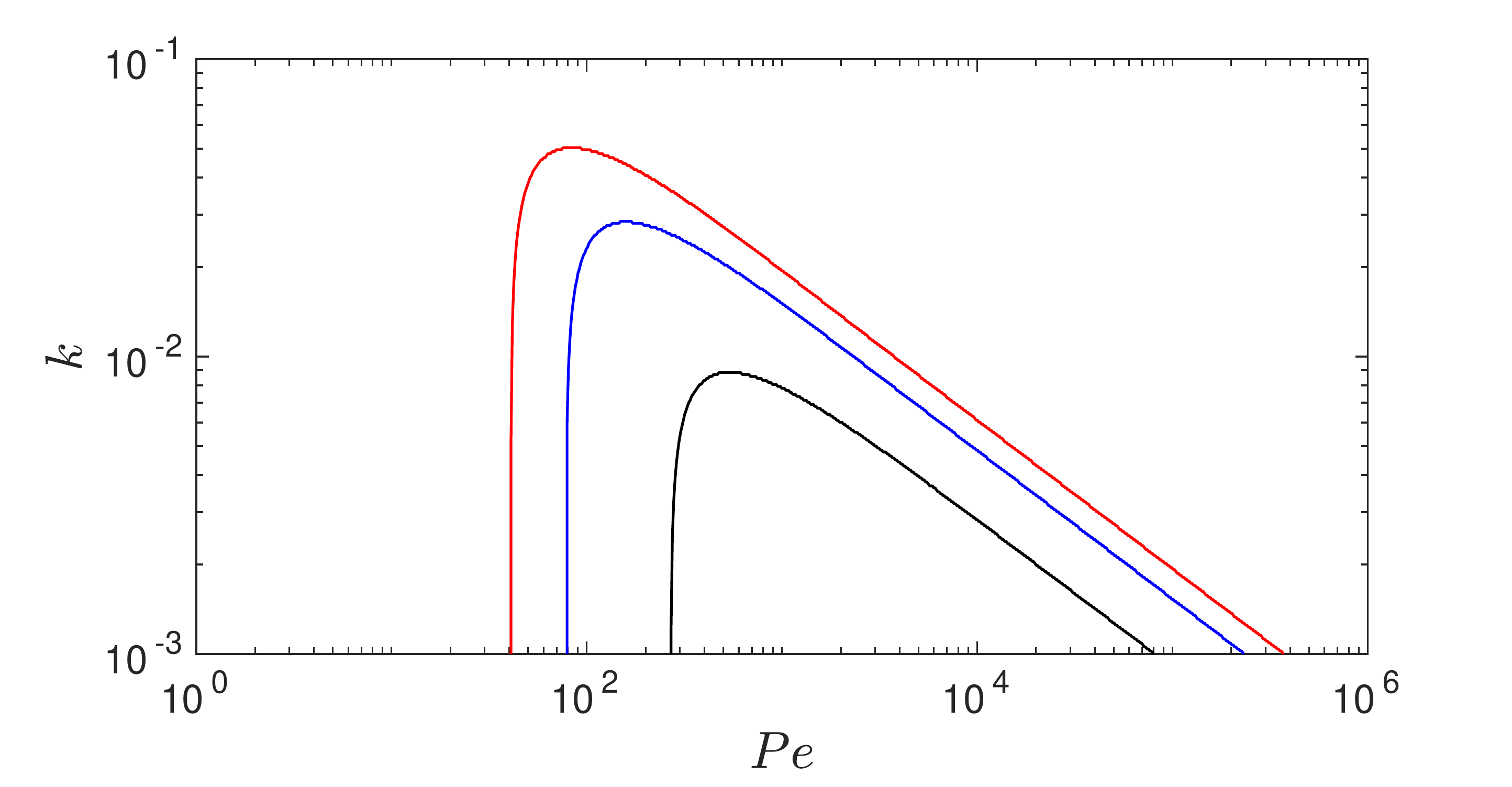}
        \caption{ Neutral stability curves of the surface mode in the $(k,\Pec)$ plane for $\alpha=0.1$ (black), $0.3$ (blue), and $0.5$ (red). Regions below the curves correspond to unstable modes.
 }
        \label{fig:7}
\end{figure}
We present the neutral stability curves in the $(\Pec, k)$ plane for three representative values of the stratification parameter, $\alpha = 0.1$, $0.3$, and $0.5$, as shown in figure~\ref{fig:7}. Each curve represents the boundary between stable and unstable regions, thereby identifying the range of wavenumbers and P\'{e}clet numbers for which the growth rate changes sign. With increasing $\alpha$, the unstable region expands in both the $k$ direction and the P\'{e}clet number direction, with the neutral curves extending toward both smaller and larger values of $\Pec$. In particular, stronger viscosity stratification reduces the critical P\'{e}clet number for instability onset, indicating that even relatively weak advective transport can sustain destabilizing viscosity gradients. The upper bound of the unstable window shifts to larger values of $\Pec$, indicating that stronger stratification enables instability to persist under increasingly advective conditions. 
The restabilization at large $\Pec$ is evident from the 
neutral curves in figure~\ref{fig:7}, where the unstable 
region is bounded from above in $\Pec$ for all values of 
$\alpha$ examined.

\begin{figure}[h]
    \centering
        \includegraphics[width=0.8\linewidth]{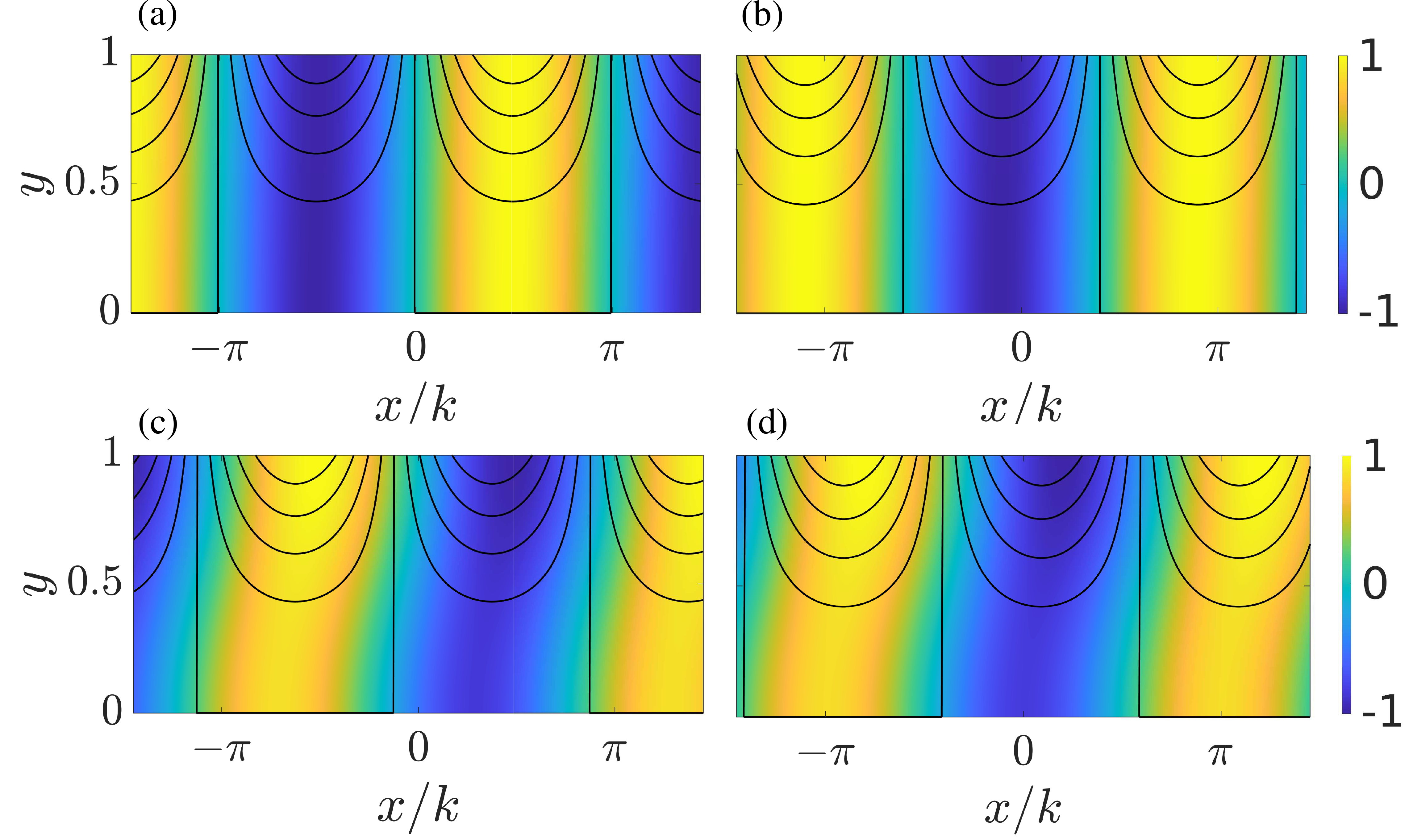}
    \caption{Perturbed viscosity field (color contours) overlaid with streamfunction perturbations (black contour lines) for $k = 0.001$, shown for different combinations of the stratification parameter $\alpha$ and Péclet number $\Pec$: (a) $\alpha = 0.3$, $\Pec = 100$; (b) $\alpha = 0.5$, $\Pec = 100$; (c) $\alpha = 0.3$, $\Pec = 1000$; (d) $\alpha = 0.5$, $\Pec = 1000$.}

    \label{fig:8}
\end{figure}
\begin{figure}[h]
    \centering
        \includegraphics[width=0.8\linewidth]{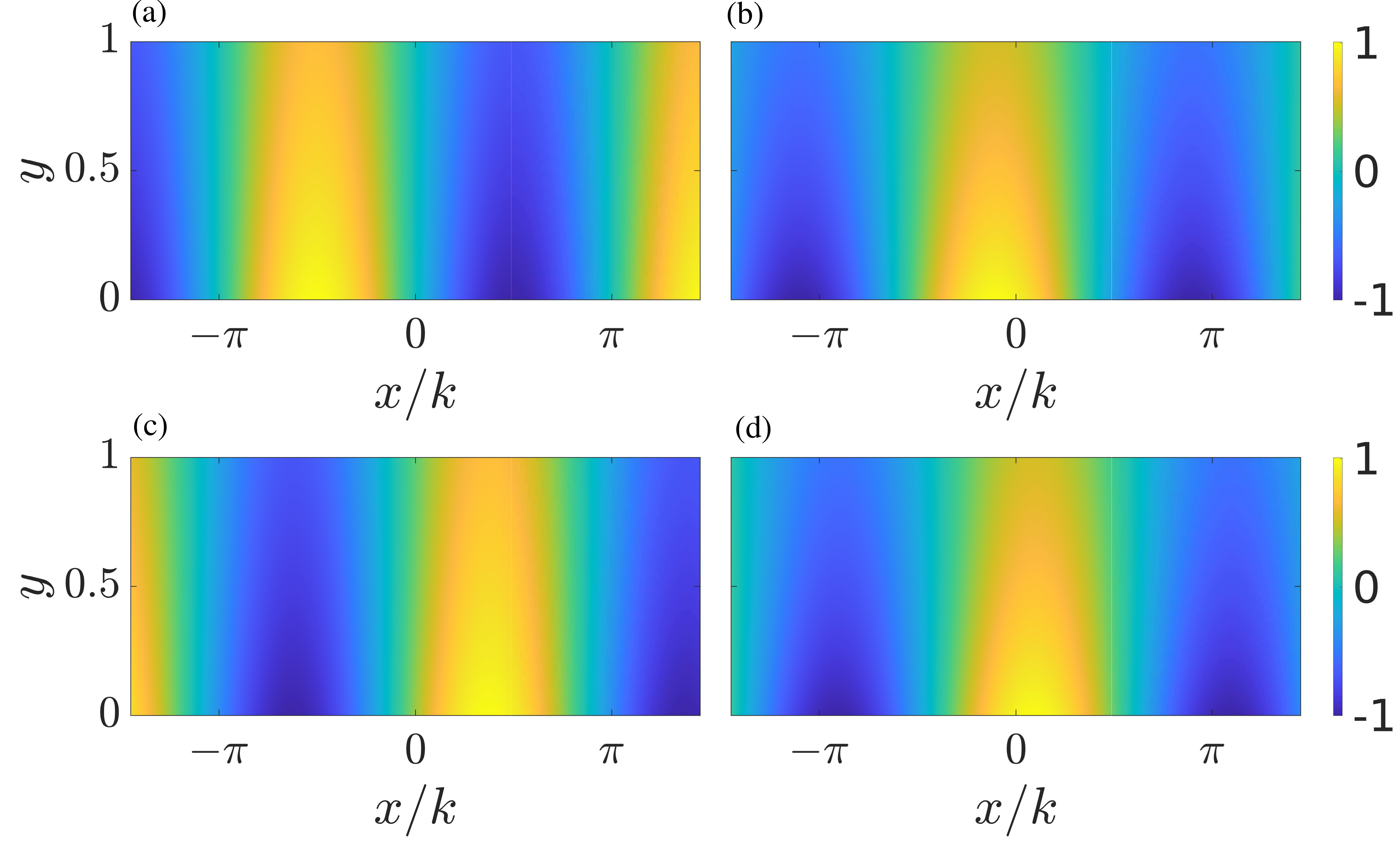}
      \caption{Vorticity field (color contours)  for \(k = 0.001\) corresponding to different combinations of stratification parameter \(\alpha\) and Péclet number \(\mathrm{Pe}\): (a) $\alpha = 0.3$, $\Pec = 100$; (b) $\alpha = 0.5$, $\Pec = 100$; (c) $\alpha = 0.3$, $\Pec = 1000$; (d) $\alpha = 0.5$, $\Pec = 1000$.}
    \label{fig:9}
\end{figure}

Figure \ref{fig:8} illustrates the structure of the unstable surface mode in a viscosity-stratified falling film, where the perturbation viscosity field is shown using color contours and the corresponding streamfunction perturbations are overlaid as black contour lines. The x-axis represents the streamwise direction of the perturbation, while the y-axis corresponds to the film thickness. The figure consists of four panels comparing different 
combinations of the stratification parameter $\alpha$ and 
P\'{e}clet number $\Pec$.
A key feature evident in the figure is the prominent tilt in the viscosity perturbation contours, indicating a clear phase shift between the viscosity field and the streamfunction. This tilt arises due to the advection of viscosity disturbances by the base flow, which is faster near the free surface. As a result, perturbations in viscosity generated by vertical displacements are transported downstream, creating a slanted pattern in the perturbation field. The strength and orientation of this tilt vary with both the stratification parameter and the P\'{e}clet number, as stronger stratification and higher P\'{e}clet numbers enhance the alignment of the perturbation with the flow direction.
In the top-left panel (low $\alpha$, low $\Pec$), the regions of maximum positive and negative viscosity perturbations are spatially aligned with the centers of the vortical structures in the streamfunction field. This indicates a nearly in-phase relationship between the viscosity and velocity perturbations, consistent with the dominance of diffusion and weak advection.
As we move to the top-right panel (low $\alpha$, high $\Pec$), a noticeable streamwise shift appears between the locations of peak viscosity perturbation and the centers of the streamfunction vortices. This shift marks the onset of a phase lag.
In the bottom-left panel (high $\alpha$, low $\Pec$), although advection remains weak, the strong background viscosity gradient results in more localized and intense perturbation structures. 
In the bottom-right panel (high $\alpha$, high $\Pec$), a strong phase shift is evident; the peaks of the viscosity perturbations are significantly displaced downstream relative to the streamfunction vortices. This spatial lag demonstrates the combined influence of strong advection and sharp viscosity gradients. The physical origin of this progressive tilt and its role in the instability mechanism are discussed in \S\ref{sec:mechanism}.
Figure~\ref{fig:9} presents the spatial distribution of the vorticity field for the same set of parameters shown in Figure~\ref{fig:8}. The panels display the streamwise variation of the vorticity field across the film thickness for different combinations of the stratification parameter and the Péclet number. As in the previous figure, the x-axis represents the streamwise coordinate, and the y-axis denotes the wall-normal coordinate.

A key observation from figure~\ref{fig:9} is the presence of a distinct phase shift in the vorticity field relative to the perturbation viscosity field shown in figure~\ref{fig:8}. For  each combination of $\Pec$ and $\alpha$, the vorticity maxima are concentrated near the wall, where the base shear $u_b' = 3(1-y)$ is largest, and are shifted in the streamwise direction relative to the corresponding viscosity perturbations. The relationship between this vorticity distribution and the instability is analyzed in \S\ref{sec:mechanism} through the vorticity-interface phase framework.

\color{black}\section{Mechanism of Instability}\label{sec:mechanism}
 
We now trace the causal chain by which viscosity stratification destabilizes the surface mode, following the vorticity-interface phase framework introduced by Hinch \cite{hinch1984} and applied to falling-film instabilities by Kelly et al. \cite{kelly1989}.
 
\subsection{Vorticity-interface phase relation}
 

Consider a sinusoidal perturbation of the free surface,
\begin{gather}
\hat{\eta} = E\sin\theta,
\end{gather}
where $\theta = k(x - c_r t)$ is the phase and $E = e^{kc_i t}$ represents the growth factor. A crest corresponds to $\theta = \pi/2$, while a trough corresponds to $\theta = -\pi/2$. We then define the perturbation vorticity as $\hat{\omega} = \partial \hat{u}/\partial y - \partial \hat{v}/\partial x$, with positive values corresponding to clockwise rotation. In terms of the streamfunction, this gives  $\hat{\omega} = (\mathcal{D}^2 - k^2)\hat{\psi}$. Using the linearised tangential stress condition together with the kinematic boundary condition at $y = 1$, the perturbation vorticity evaluated at the free surface can be written as
\begin{equation}\label{eq:omega_fs}
\omega\big|_{y=1} = -2k^2 c_i\cos\theta + \left[\frac{3}{\mu_b(1)} - 2k^2(c_r - u_b(1))\right]\sin\theta.
\end{equation}
The above expression takes the compact form $\omega \propto \sin(\theta + \phi)$, where
\begin{equation}\label{eq:tanphi}
\tan\phi =
\frac{-2k^2 c_i}{\dfrac{3}{\mu_b(1)} - 2k^2(c_r - u_b(1))}.
\end{equation}

\begin{figure}[h]
    \centering
        \includegraphics[width=0.8\linewidth]{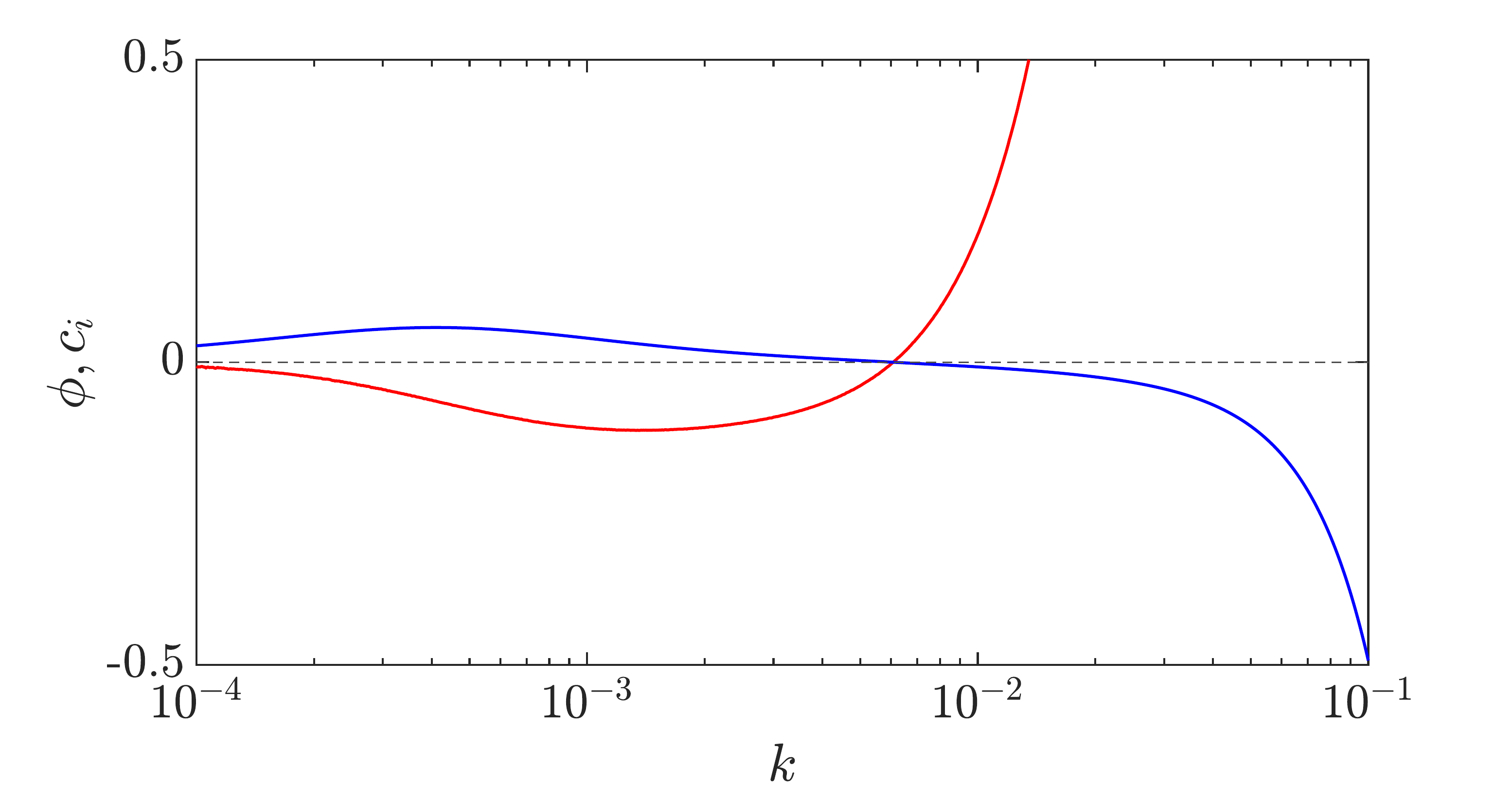}
      \caption{Variation of the vorticity-interface phase angle $\phi$ (red) and the growth rate $c_i$ (blue) with  wavenumber $k$ for $\mathrm{Pe} = 10^4$, $\alpha = 0.5$, 
    and $\theta = 45^\circ$. }
    \label{fig:10}
\end{figure}

\begin{figure}[ht]
    \centering
    \begin{subfigure}[b]{0.45\textwidth}
        \centering
        \includegraphics[width=\textwidth]{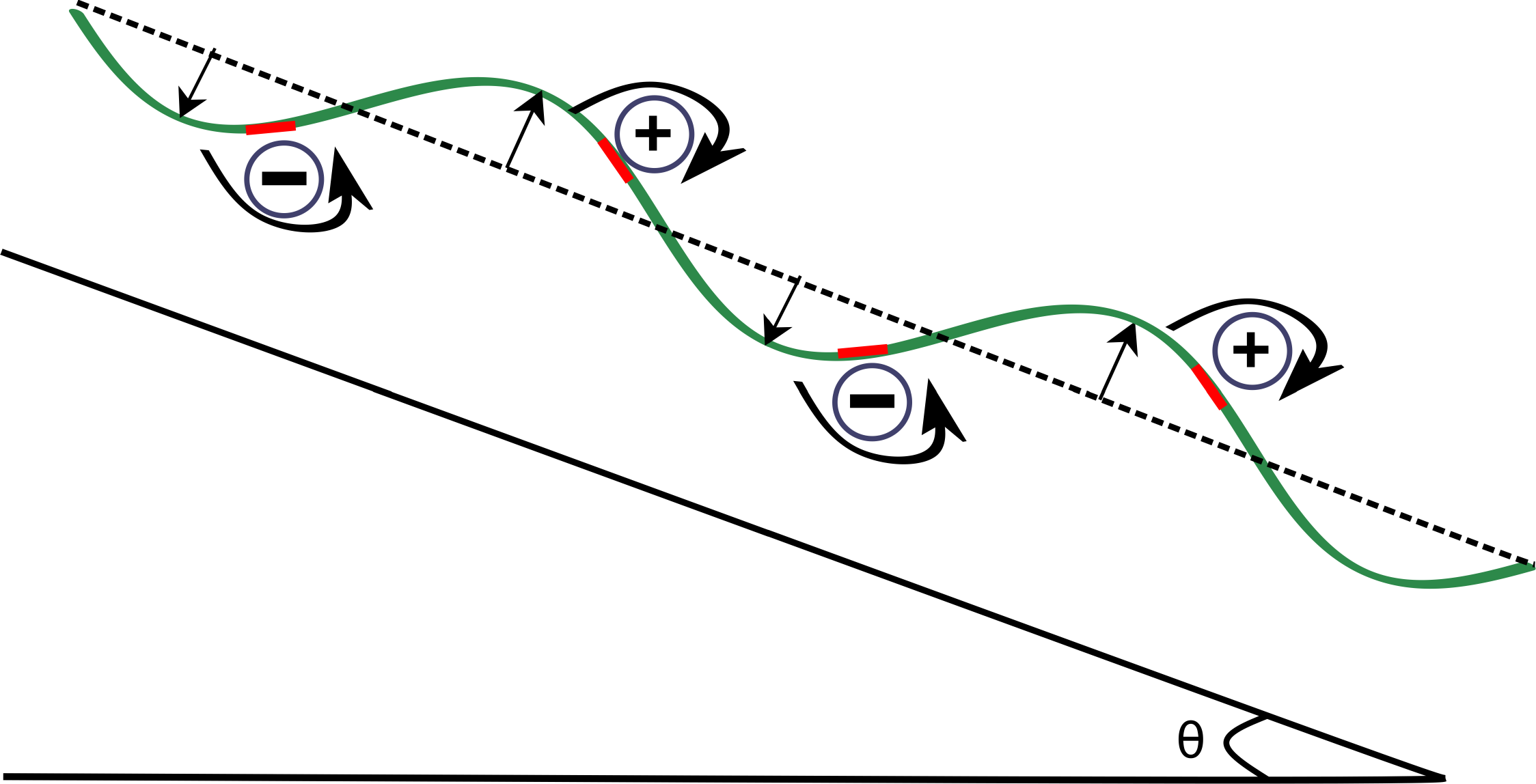}
        \label{fig:mechanism_unstable}
    \end{subfigure}
    \hfill
    \begin{subfigure}[b]{0.45\textwidth}
        \centering
        \includegraphics[width=\textwidth]{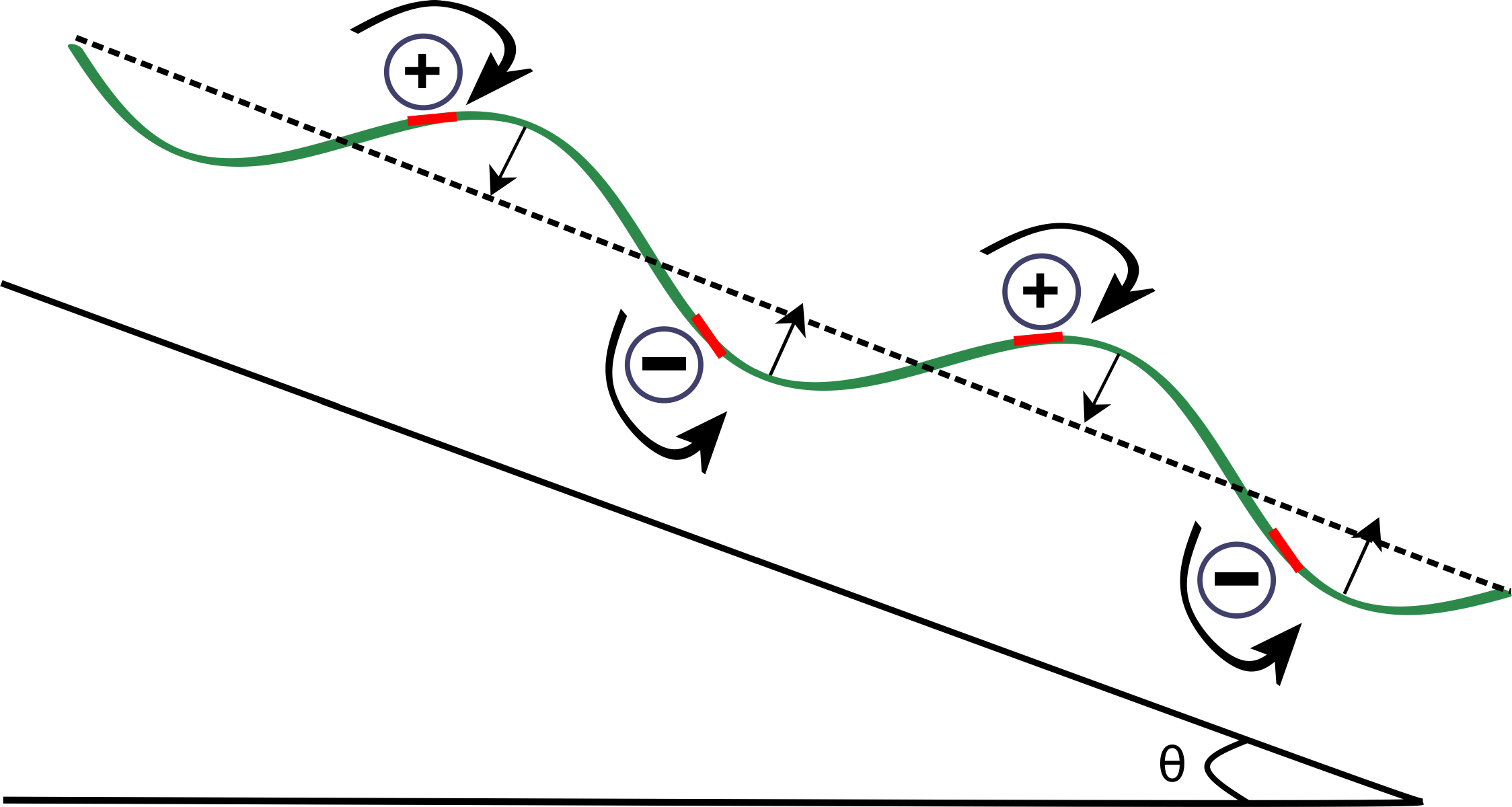}
        \label{fig:mechanism_stable}
    \end{subfigure}
    \caption{Mechanism of the inertialess instability. 
    (a)~$c_i>0$
    (b)~$c_i<0$}
    \label{fig:mechanism}
\end{figure}

Since we have a long-wave instability in the system, $3/\mu_b(1) \gg 2k^2|c_r - u_b(1)|$. This implies that the denominator is strictly positive, leaving the sign of $\phi$ to be dictated by $c_i$. When $c_i > 0$, $\phi < 0$, the vorticity lags the interface. The lagging vorticity drives fluid upward on the upstream side of crests and downward upstream of troughs, reinforcing the displacement and producing instability (figure \ref{fig:mechanism}a). When $c_i < 0$, the vorticity leads ($\phi > 0$) and opposes the displacement (figure \ref{fig:mechanism}b). Figure \ref{fig:10} shows direct numerical confirmation of this phase relationship. The phase angle $\phi$, computed from the eigenfunctions at each wavenumber, is negative precisely where $c_i > 0$ and is in exact agreement with equation \eqref{eq:tanphi}. At $c_i = 0$ the vorticity is in phase with the interface and the film is neutrally stable.


\subsection{How viscosity stratification creates the lag}\label{sec:feedback mechanism} 

While we have shown that the vorticity-interphase phase difference drives the instability, it remains unclear how viscosity stratification generates it. To understand this, we turn back to the long-wave expansion of~\S\ref{sec:longwave} and identify a two-step feedback loop.
 
 
\medskip
\noindent\textit{Step 1: Generation of $\hat{\mu}_1$ in $90^\circ$ out of phase.}
At leading order, a sinusoidal interface deflection $\hat{\eta}$ induces a real perturbation streamfunction $\hat{\psi}_0(y)$. This streamfunction advects the base-state viscosity gradient $\mu_b'$ through the source term $-\mu_b'\hat{\psi}_0$ in the $\mathcal{O}(k)$ viscosity equation
\begin{equation}\label{eq:mu1_source}
\mathcal{D}^2\hat{\mu}_1
= i\,\Pec\bigl[(u_b - c_0)\hat{\mu}_0 - \mu_b'\hat{\psi}_0\bigr].
\end{equation}
The crucial observation is the factor of $i$ on the right-hand side. Since $\hat{\psi}_0$ and $\hat{\mu}_0$ are both real, the resulting $\hat{\mu}_1$ is purely imaginary and $90^\circ$ out of phase with $\hat{\psi}_0$. Physically, the perturbation streamfunction displaces fluid parcels vertically. The base-state viscosity gradient converts these displacements into viscosity anomalies, and the advection-diffusion balance tilts these anomalies downstream, producing a viscosity perturbation that is phase-shifted relative to the interface. The differential base flow $u_b(y)$, which is faster near the free surface, further tilts the $\hat{\mu}_1$ field, as evident in the eigenfunction contour plots of figure~\ref{fig:8}.

 
\medskip
\noindent\textit{Step 2: Vorticity feedback via the momentum equation.}
The phase-shifted viscosity perturbation $\hat{\mu}_1$ enters the streamfunction equation \eqref{eqn:mom_vis} through the forcing term $(\mathcal{D}^2 + k^2 (u_b'\hat{\mu})$. Because $\hat{\mu}_1$ is $90^\circ$ out of phase with $\hat{\psi}_0$, this forcing acts as a perturbation vorticity source that is no longer aligned with the original interface deflection. The resulting correction to the streamfunction at $\mathcal{O}(k)$ yields $c_{1i} > 0$, leading to the vorticity being displaced into the lagging configuration, and the instability is triggered.

\medskip
In the unstratified limit ($\alpha = 0$), the source term 
$-\mu_b'\hat{\psi}$ vanishes, the feedback loop is inactive, and the surface mode is stable.

 

\subsection{The finite-$\Pec$ window}\label{sec:mechanism-Pe}

A distinctive feature of the instability is its confinement to a finite window of P\'{e}clet numbers (figures \ref{fig:3} and \ref{fig:7}). We posit that the role of $\Pec$ is to control the efficiency of the phase-shifting mechanism.
 
\medskip
\noindent\textit{Low $\Pec$ (diffusion-dominated):}
When $\Pec$ is small, diffusion rapidly homogenizes the viscosity perturbation. The source term in \eqref{eq:mu1_source} generates $\hat{\mu}_1 \propto i\,\Pec$. Thus, the $90^\circ$ out of phase component is small and vanishes as $\Pec \to 0$, and the feedback loop is effectively switched off. The flow therefore remains stable since the viscosity-stratification coupling is not sufficient enough to to shift the vorticity into the lagging configuration.



 
\medskip

\noindent\textit{Intermediate $\Pec$:}
At moderate $\Pec$, the viscosity perturbation is large enough to provide effective feedback on the vorticity field. Thus, the growth rate attains its peak, as observed in the contour plots in figure \ref{fig:3}.

\medskip
\noindent\textit{Large $\Pec$ (advection-dominated):}
When $\Pec$ is large, the advection dominates over the diffusion. In this limit, the viscosity perturbation approaches the frozen-field solution $\hat{\mu} \approx \mu_b'\hat{\psi}/(u_b - c)$, which is
real-valued and therefore in phase with the perturbation streamfunction. The component of $\hat{\mu}$ that is essential for the vorticity feedback (Step 2) thus vanishes, deactivating the phase-shifting mechanism and rendering the system stable. The numerical neutral curves (figure~\ref{fig:7}) confirm this stabilization at large $\Pec$ for all values of $\alpha$ examined.


 
\medskip
\noindent
This finite-$\Pec$-window behavior is structurally identical to the surfactant-driven Marangoni instability in creeping two-layer flows \citep{frenkel2002,wei2005}, where the surfactant concentration plays the role of the viscosity perturbation, and Marangoni stresses replace the viscosity-induced vorticity feedback. In both systems, the transported scalar must be coupled to the momentum equation through a stress modification, and the instability arises only when the scalar transport is neither too diffusive nor too advective.

\color{black}
\subsection{Necessity of both $\mu_b'$ and $\hat{\mu}$}
 
Finally, to confirm that the two-way coupling is essential, we consider four limiting cases by selectively retaining or suppressing the base-state viscosity gradient $\mu_b'$ and the viscosity perturbation $\hat{\mu}$, as summarised in table \ref{tab:cases}.
 
\begin{table}[t]
\centering
\begin{tabular}{clccl}
\hline
Case & Description & $\mu'_b$ & $\hat{\mu}$ & Stability \\
\hline
1 & Fully coupled viscosity-stratified system & $\neq 0$ & Present & Unstable \\
2 & Uniform viscosity (no stratification, no transport)     & $= 0$    & Absent  & Stable \\
3 & Uniform base state with viscosity transport & $= 0$ & Present & Stable \\
4 & Stratified base state without viscosity perturbations & $\neq 0$ & Absent & Stable \\
\hline
\end{tabular}
\caption{Role of the base-state viscosity gradient $\mu'_b$ and
viscosity perturbations $\hat{\mu}$ in the instability.}
\label{tab:cases}
\end{table}
 
In Case 1, both $\mu_b' \neq 0$ and $\hat{\mu}$ are retained. This is the full problem, which exhibits the instability described above. 
In Case 2, the viscosity is uniform ($\mu_b' = 0$), and no viscosity perturbations arise. This recovers the classical unstratified falling film, which is stable in the Stokes limit. 
In Case 3, the base-state viscosity is uniform ($\mu_b' = 0$), but the viscosity perturbation equation is retained. Since the source term $-\mu_b'\hat{\psi}$ in equation \eqref{eq:par1_vis} vanishes when $\mu_b' = 0$, no viscosity perturbation is generated by the flow, and the system remains stable. 
In Case 4, the base-state viscosity is stratified ($\mu_b' \neq 0$) but the viscosity perturbation is suppressed ($\hat{\mu} = 0$). The momentum equation \eqref{eqn:mom_vis} then reduces to the homogeneous variable-coefficient problem without viscosity-perturbation feedback, and the flow is again stable.
 
The instability therefore requires both links of the feedback loop: the base-state gradient $\mu_b'$ provides the source that generates $\hat{\mu}$ from the perturbation flow (Step 1), while the resulting $\hat{\mu}$ feeds back on the momentum equation through the forcing $(\mathcal{D}^2 + k^2)(u_b'\hat{\mu})$ (Step 2), shifting the vorticity into the lagging configuration that drives the instability. Removing either link breaks the loop and restores stability.

We further note that the instability persists even when the base-state viscosity gradient $\mu_b'$ is removed from the left-hand side of the momentum equation \eqref{eqn:mom_vis}, i.e., when the terms $-2\mu_b'(\mathcal{D}^2 - k^2)\mathcal{D}$ and $-\mu_b''(\mathcal{D}^2 + k^2)$ are set to zero. In this case, the momentum operator reduces to $-(\mathcal{D}^2 - k^2)^2$, while $\mu_b'$ is retained in the transport equation \eqref{eq:par1_vis} and the viscosity-perturbation feedback on the right-hand side of \eqref{eqn:mom_vis} is kept intact. This confirms that the variable-coefficient structure of the momentum operator is not an essential ingredient of the instability, and that the destabilization is driven entirely by the two-way perturbation coupling identified in Steps 1 and 2 in \S \ref{sec:feedback mechanism}.

\section{Conclusions}

We have demonstrated that the mere introduction of viscosity stratification is sufficient to trigger a surface-mode instability in a gravity-driven falling film, even in the absence of fluid inertia, a regime where the classical Kapitza instability is absent. This instability was observed to arise solely from perturbation-level coupling between the viscosity and velocity fields, while the base-state velocity profile was taken to be the Nusselt solution, unaltered by the stratification of viscosity.

 
The instability mechanism, elucidated through the vorticity-interface phase framework of Hinch~\cite{hinch1984} and Kelly et
al. \cite{kelly1989}, operates through a two-step feedback loop. The perturbation streamfunction, induced by the interface deflection, advects the base-state viscosity gradient to produce a viscosity perturbation $\hat{\mu}_1$ that is $90^\circ$ out of phase with $\hat{\psi}_0$. This phase-shifted $\hat{\mu}_1$ feeds back on the momentum equation as a vorticity source, displacing the surface vorticity into the lagging configuration that amplifies the interface deformation.
 
Neutral stability maps in the $(\alpha, \Pec)$ and $(k, \Pec)$ planes reveal that instability is confined to a finite window of P\'{e}clet numbers, bounded below by diffusive smoothing and above by advective decorrelation. Increasing the stratification parameter~$\alpha$ widens this window, reduces the critical~$\Pec$ for onset, broadens the range of unstable wavenumbers, and increases the growth rate. The long-wave analytical results at $\mathcal{O}(k)$ are in good agreement with the numerical solutions obtained via Chebyshev spectral collocation.

The finite-$\Pec$-window structure and the underlying phase-shift mechanism place the present instability within a broader class of scalar-mediated, inertialess instabilities in thin-film flows. In the surfactant-driven Marangoni instability of creeping two-layer flows \citep{frenkel2002,wei2005}, an interfacial surfactant concentration couples to the momentum equation through Marangoni stresses; in the thermocapillary instability of a film overlying a phase boundary \citep{dhasetal2024jfm}, the temperature field couples through surface-tension gradients at the free surface. In each case, a transported scalar-surfactant concentration, temperature, or, as shown here, viscosity generates a phase-shifted perturbation that feeds back on the flow, and the instability exists only when the scalar transport is neither too diffusive nor too advective. The present work extends this class from interfacial scalars to bulk viscosity stratification, demonstrating that the coupling need not be confined to the free surface to produce an inertialess instability.
 
The present analysis adopts a linear, prescribed viscosity profile with constant diffusivity $D_s$, considers only two-dimensional perturbations, and assumes the base-state velocity is unaffected by the stratification. Under these assumptions, the analysis isolates viscosity stratification as a sufficient condition for triggering a surface-mode instability in the inertialess limit. In particle-laden films, however, the viscosity distribution evolves dynamically through shear-induced migration, the base flow is modified by the resulting concentration profile \citep{timberlake2005,dhas2022}, and particle-phase normal stresses contribute additional momentum coupling. Extending the present framework to include these effects, along with nonlinear viscosity-concentration relationships relevant to dense suspensions, would clarify the extent to which the mechanism identified here persists in such systems and what additional physics may modify or enhance it. Nevertheless, the present results establish that viscosity stratification constitutes a minimal and sufficient mechanism for inertialess surface-mode instability in falling films.

\color{black}

\section*{Acknowledgments}
The authors acknowledge the financial support received from the Science and Engineering Research Board (SERB), Government of India, through the Core Research Grant (CRG/2023/008504). The authors also thank the Indian Institute of Technology Madras for providing the computational and infrastructural facilities that supported this work.

\appendix

\section{Integrals arising in the long wave analysis}\label{appA}
Here, the expressions for $a_1$, $a_2$, and $a_3$ used in the dispersion relation \eqref{eqn:c0} are presented here.

\begin{gather}
    a_1=1,\\
    a_2= -(\mathcal{I}_1(1)+2u_b(1)+\mathcal{J}_0(1)),\\
    a_3= u_b(1)^2+u_b(1)\mathcal{I}_1(1)+u_b(1)\mathcal{J}_0(1)+\mathcal{I}_1(1)\mathcal{J}_0(1)-\mathcal{I}_2(1)\mathcal{J}_1(1).
\end{gather}
where 
\begin{eqnarray}
&&\mathcal{I}_1(y)=\int_0^y dz_1 \int_0^{z_1} dz_2 \,\frac{3}{\mu_b}\\
&&\mathcal{I}_2(y)=\int_0^y dz_1 \int_0^{z_1} dz_2 \frac{u_b'}{\mu_b}
\end{eqnarray}
\begin{eqnarray}
    \mathcal{J}_0(y)&=&\int_0^y \left\{u^*(z)+\mu'_b\,\mathcal{I}_2(z)\right\} dz\\
     \mathcal{J}_1(y)&=&\int_0^y \mu'_b(z)\,\mathcal{I}_1(z) dz
\end{eqnarray}
\begin{equation}
    \mathcal{G}_{0}\left( y\right) =\int ^{y}_{0}\mathcal{J}_{0}\left( z\right) dz-\int ^{y}_{0}c^{\ast }zdz
\end{equation}
\begin{equation}
    \mathcal{G}_{1}\left( y\right) =\int ^{y}_{0}\mathcal{J}_{1}\left( z\right) dz
\end{equation}

\begin{eqnarray}
   \mathcal{H}_{0}(y)= \int_{0}^{y} dz_{1} \int_{0}^{z_{1}} \frac{1}{\mu_{b}}u_b'\mathcal{G}_{0}\left( z_2\right)dz_{2} \\
   \mathcal{H}_{1}(y)=\int_{0}^{y} dz_{1} \int_{0}^{z_{1}} \frac{1}{\mu_{b}}u_b'\mathcal{G}_{1}\left( z_2\right)dz_{2} \\
   \mathcal{H}_{2}(y)=\int_{0}^{y} dz_{1} \int_{0}^{z_{1}} \frac{z_2}{\mu_{b}} dz_{2} \\
   \mathcal{H}_{3}(y)=\int_{0}^{y} dz_{1} \int_{0}^{z_{1}} \frac{1}{\mu_{b}} dz_{2} 
\end{eqnarray}

\begin{eqnarray}
    \mathcal{L}_{0}(y)= \int_0^y \left[(u^*-c^*) \mathcal{G}_{0}\left( z\right)+\mu'_b  \mathcal{H}_{0}(z)\right] dz \\
    \mathcal{L}_{1}(y)= \int_0^y \left[ (u^*-c^*) \mathcal{G}_{1}\left( z\right)+ \mu'_b  \mathcal{H}_{1}(z)\right] dz \\
    \mathcal{L}_{2}(y)= \int_0^y \mu'_b \mathcal{H}_{2}(z) dz\\
    \mathcal{L}_{3}(y)= \int_0^y \mu'_b \mathcal{H}_{3}(z) dz
\end{eqnarray}
\begin{align}
c_{1} =
\frac{1}{
\left[
1 + \frac{1}{c^*}
\frac{\mathcal{I}_2(1)}{c^* - \mathcal{I}_1(1)}
\left(
1 - \frac{\mathcal{I}_1(1)}{c^* - \mathcal{I}_1(1)}
\right)
\mathcal{J}_1(1)
\right]
}
\Bigg(
i\,\Pec\,\mathcal{L}_0(1)
- i\,\Pec\,\frac{\mathcal{I}_2(1)}{\mathcal{I}_1(1) - c^*}\,\mathcal{L}_1(1)
- i\,\mathcal{F}\,\frac{\mathcal{I}_2(1)}{\mathcal{I}_1(1) - c^*}\,\mathcal{L}_2(1)
\nonumber \\
- \frac{u_b'(1)\hat{\mu}_1(1)}{\hat{\mu}_0}\,\mathcal{L}_3(1)
+ \Bigg[
\frac{i\,\mathcal{F}}{3}\,\frac{\mathcal{I}_2(1)}{\mathcal{I}_1(1) - c^*}
- i\,\Pec \left(
\frac{\mathcal{H}_0(1)}{\mathcal{I}_1(1) - c^*}
- \frac{\mathcal{I}_2(1)\mathcal{H}_1(1)}{\left(\mathcal{I}_1(1) - c^*\right)^2}
\right)
\nonumber \\
+ i\,\mathcal{F}\,\frac{\mathcal{I}_2(1)}{\left(\mathcal{I}_1(1) - c^*\right)^2}
\left(
\mathcal{H}_2(1) - \frac{\mathcal{I}_1(1)}{3}
\right)
- \frac{u_b'(1)\hat{\mu}_1(1)}{\hat{\mu}_0}
\left(
\frac{\mathcal{H}_3(1)}{c^* - \mathcal{I}_1(1)}
\right)
\Bigg]\mathcal{J}_1(1)
- \frac{i}{\Pec}
\Bigg)
\label{eqn:c1}
\end{align}
where $c^*=c_0-u_b(1)$ and $u^*=u_b(y)-u_b(1)$.



\printcredits

\bibliographystyle{unsrt}
\bibliography{biblio.bib}

\end{document}